\documentclass[12pt]{iopart}

\usepackage{graphicx}
\usepackage[usenames]{color}
\newcommand{\beq}{\begin{equation}}
\newcommand{\eeq}{\end{equation}}
\newcommand{\beqa}{\begin{eqnarray}}
\newcommand{\eeqa}{\end{eqnarray}}
\newcommand{\ba}{\begin{array}}
\newcommand{\ea}{\end{array}}

\begin{document}

\title{Universal thermodynamics of a strongly interacting Fermi gas: theory
versus experiment}

\author{Hui Hu$^{1,2}$\footnote{hhu@swin.edu.au}, Xia-Ji Liu$^{1}$, and Peter D. Drummond$^{1}$}

\address{$^{1}$\ ARC Centre of Excellence for Quantum-Atom Optics, Centre
for Atom Optics and Ultrafast Spectroscopy, Swinburne University of Technology, Melbourne 3122, Australia
\\
 $^{2}$\ Department of Physics, Renmin University of China, Beijing
100872, China}

\begin{abstract} 
Strongly interacting, dilute Fermi gases exhibit a scale-invariant,
universal thermodynamic behaviour. This is notoriously difficult to
understand theoretically because of the absence of a small interaction
parameter. Here we present a systematic comparison of theoretical
predictions from different quantum many-body theories with recent
experimental data of Nascimbène \textit{et. al.} {[}Nature \textbf{463},
1057 (2010){]}. Our comparisons have no adjustable parameters, either
theoretically or experimentally. All the model approximations seem
to be fluctuating around and not converging towards the experimental
data. It turns out that a simple Gaussian pair fluctuation theory gives 
the best quantitative agreement, except at the critical superfluid
transition region. In the normal state, we also calculate the equation
of state by using a quantum cluster expansion theory and explore in
detail its applicability to low temperatures. Using the accurate experimental
result for the thermodynamic function $S(T)$, we determine the temperature
$T$ of a trapped Fermi gas at unitarity as a function of a non-interacting
temperature $T_{i}$ which can be obtained by an adiabatic sweep to
the free gas limit. By analyzing the recent experimental data, we
find a characteristic temperature $(T/T_{F})_{0}=0.19\pm0.02$
or $(T_{i}/T_{F})_{0}=0.16\pm0.02$ in a harmonic trap, below which there are
deviations from normal Fermi-liquid-like behavior that may be attributed to 
pairing effects.  Here $T_{F}$ is the Fermi temperature for a trapped ideal, 
non-interacting Fermi gas. Our thorough comparison may shed light on 
further theoretical development of strongly interacting fermions.
\end{abstract}


\maketitle

\section{Introduction}

\label{I}

The recent discovery of broad Feshbach resonances in two-component
atomic Fermi gases has opened a new era in the study of strongly interacting
fermions \cite{review,cm1,cm2,cm3,jila1,mit1,mit2}. By tuning an
external magnetic field across the Feshbach resonance, the interatomic
attractions can be changed precisely from weak to infinitely strong,
leading to the observation of a crossover from a Bardeen-Cooper-Schrieffer
(BCS) superfluid to a Bose-Einstein condensation (BEC). At resonance,
the \textit{s}-wave scattering length $a_{s}$ diverges ($a_{s}=\pm\infty$)
and the two-body scattering amplitude reaches the maximum value allowed
by quantum mechanics due to unitarity. Many unique properties are
anticipated in this strongly interacting limit, including a high superfluid
transition temperature and an exotic normal state with a pseudogap.

Most interesting of all is fermionic universality. This means that
all strongly interacting, dilute Fermi gases behave identically, regardless
of the details of the interaction. Their properties depend only on
temperature, together with a scaling factor equal to the average particle
separation \cite{heiselberg,ho1,thomas,natphys}. This limit promises
to bring a new rigour and simplicity to the understanding of strongly
correlated Fermi gases. Because of universality, it should be feasible
to understand other strongly interacting Fermi superfluids from experiments
in the highly controlled environment of an atomic physics laboratory.
Possible examples include neutron stars and high-$T_{c}$ superconductors.

Intense experimental investigations have been carried out to understand
fermionic universality, in particular, its implication to the thermodynamic
properties \cite{duke1,duke2,duke3,duke4,rice,jila2,tokyo,ens,ens2}. Pioneering
observations were carried out at Duke University \cite{duke1,duke2,duke3,duke4},
which made the first attempt to reach the unitarity limit in $^{6}$Li
gas in 2002 \cite{duke1}. The stability of atomic Fermi gases with
strongly attractive interactions was observed. The ground state energy
was found to be reduced significantly compared to its ideal, non-interacting
limit. The reduction factor $\beta$ (or $\xi=1+\beta$) has now been
determined accurately to within a few percent: $\beta\simeq-0.59\pm0.01$,
after substantial experimental effort \cite{ens2}.

In this paper we show that recent highly accurate measurements \cite{ens}
on strongly interacting $^{6}$Li give the most stringent test to
date of fermionic strongly coupling many-body theories. In fact, these
experiments determine the whole set of universal thermodynamic functions
for a trapped Fermi gas at unitarity. As well as measuring bulk thermodynamic
properties, the data can be used to determine the energy and entropy,
$E(T)$ and $S(T)$, of a trapped gas. Due to the use of larger samples,
the accuracy is even better than that achieved at Duke. With unprecedented
precision, these new universal functions therefore provide an unbiased
test of theoretical predictions. This comparison, without any fitting
parameters, indicates that while a BCS-type mean field theory is certainly
incorrect, an extension using a simple Gaussian pair fluctuation theory
provides overall the best agreement with experiment, except at the
critical superfluid transition region.

To give some background to these developments, the first energy and
heat capacity measurements as a function of temperature were performed
by Kinast \textit{et al. }\cite{duke2}. However, due to the lack
of reliable thermometry in the strongly interacting regime, an empirical
thermometry was used. Conversion of measured results to real temperature
required a particular strong-coupling theory, and was therefore model-dependent.
This difficulty of model dependence was overcome at the end of year
2006\textit{ }\cite{duke3,duke4}, by means of direct measurements
of entropy instead of temperature. In this way, both energy $E$ and
entropy $S$ were determined without invoking any specific theoretical
model. These pioneering and very important model-independent measurements
had an accuracy at the level of only a few percent.

At the same period, the potential energy of a strongly interacting
$^{40}$K gas was also measured at JILA \cite{jila2}. The temperature
was characterized in terms of the non-interacting temperature of an
adiabatically equivalent ideal Fermi gas. This is therefore equivalent
to an entropy measurement. These sets of experimental data, together
with results of another $^{6}$Li experiment at Rice, were analyzed
by the present authors \cite{natphys}. The result was that all the thermodynamic
data lay on a single universal curve. This gave the first, very strong
evidence for the universal thermodynamics of a strongly interacting
Fermi gas \cite{natphys}.

In parallel with these ground-breaking experiments, there have been
numerous theoretical studies of the thermodynamics of a strongly interacting
Fermi gas. In the absence of exact solutions, the methods used were
either strong-coupling perturbation theories \cite{nsr,sademelo,gg94,engelbrecht,ohashi,perali,chen,lh1,hld1,hld2,lh2,gg07,lhd,nishida,nikolic,veillette,diener,hld3,gg08,combescot}
or \emph{ab-initio} quantum Monte Carlo (QMC) methods \cite{carlson1,astrakharchik,bulgac1,burovski,akkineni,bulgac2,carlson2,bulgac3,burovski2,morris}.
However, a deeper understanding is made difficult by the absence of
a controllable small interaction parameter \cite{bertsch}. The use
of standard perturbation theories thus requires infinite order expansions.
These typically require truncations of sets of diagrams which cannot
be fully justified \emph{a priori}. Numerically exact QMC calculations
are very helpful and can provide unbiased benchmarks, provided that there 
is an appropriate extrapolation of the lattice results to large lattice size or,
in the diagrammatic Monte Carlo case, to zero range potentials.

The first theoretical explanation of the heat capacity at unitarity
was given by Chen \textit{et al.} \cite{duke2}, using a pseudogap
theory. In this study an empirical temperature was converted to an
approximate real temperature. The present authors subsequently gave
a theoretical prediction for the both the homogeneous and trapped
equation of state at unitarity \cite{hld2}. This used a Gaussian
pair fluctuation (GPF) theory below threshold, thus extending an approach
proposed initially by Nozières and Schmitt-Rink (NSR) \cite{nsr,hld1}
for the above threshold case. We showed that the conversion of empirical
temperature to actual temperature is strongly model-dependent.

Thus, in principle one cannot obtain accurate information about the
real temperature from these empirical temperature measurements, without
a reliable strong-coupling theory. The model independent measurement
of energy as a function of entropy, $E(S)$, by Luo \textit{et al.}
\cite{duke3,duke4} was therefore a crucial experimental advance.
This provided the first data that could be used to quantitatively
compare different strong-coupling theories without any free parameters.
One such comparison was performed by the present authors \cite{hld3},
by using different perturbation theories and available QMC results
\cite{bulgac2}. Even so, it was still impossible to determine the
dependence of the energy on temperature $E(T)$ and of the entropy
on temperature $S(T)$, due to difficulties in determining the absolute
temperature $T$. Moreover, the measurements and comparisons were
restricted to the case of a trapped Fermi gas.

Most recently, a general method was developed by Nascimbène \textit{et
al.} at ENS to measure the bulk equation of state of a homogeneous
Fermi gas of lithium-6 atoms \cite{ens}, following a theoretical
proposal by Ho and Zhou \cite{ho2}. The local pressure $P(\mu(z),T)$
or the local thermodynamic potential $\Omega(\mu(z),T)=-P(\mu(z),T)\delta V$ ($\delta V$ is the volume of a cell at the position $z$) 
of the trapped gas was directly probed using \emph{ in situ} images
of the doubly-integrated density profiles along the long $z$-axis.
The temperature was then determined by using a new thermometry approach
employing a $^{7}$Li impurity. The chemical potential could also
be determined using the local density approximation, with $\mu(z)=\mu_{0}-V_{trap}(z)$
and the central chemical potential $\mu_{0}$ being determined appropriately.
By introducing a universal function \cite{footnote} \begin{equation}
h\left[\zeta\right]=\frac{\Omega(\mu,T)}{\Omega^{(1)}(\mu,T)},\label{hzeta}\end{equation}
 experimentalists were able to determine $h(\zeta)$ with very low
noise. Here, $\zeta\equiv\exp\left(-\mu/k_{B}T\right)$, $\Omega(\mu,T)$
is the interacting thermodynamic potential and $\Omega^{(1)}(\mu,T)$
is the thermodynamic potential of an ideal two-component Fermi gas.
This precise measurement allows a direct comparison with many-body
theories developed for a uniform Fermi gas \cite{ens}.

Our main results may be summarized as follows. First, though the theoretical
predictions from all the model approximations seem to fluctuate around
the experimental data, it turns out that the simplest Gaussian
pair fluctuation theory gives the best description of the observed
thermodynamic properties, except in the vinicity of the superfluid
transition point. Second, using the measured universal functions as
a benchmark, we examine the applicability of a quantum cluster (virial)
expansion method \cite{ho3,ourve}. We find that for a trapped gas,
up to the leading interaction effect (second order), the expansion
is quantitatively reliable down to $T\simeq0.7T_{F}$. This limit
can be decreased further to $T\simeq0.4T_{F}$, with inclusion of
higher order virial coefficients (i.e., up to fourth order). Thus,
we demonstrate clearly the usefulness of quantum cluster expansion
in the study of a normal, but strongly interacting quantum gas.

Finally, we note that the temperature of a trapped Fermi gas at unitarity
is often characterized by a non-interacting temperature $T_{i}$ obtained
by an adiabatic sweep to the ideal gas limit. Using the accurate experimental
data for $S(T)$, we calculate the relation $T(T_{i})$, which shows
an apparent kink at low temperatures. We therefore determine a characteristic
temperature $(T/T_{F})_{0}=0.19\pm0.02$ or $(T_{i}/T_{F})_{0}=0.16\pm0.02$
for a trapped Fermi gas at unitarity, below which the thermodynamic functions start 
to deviate from normal Fermi-liquid-like behavior due to pairing effects.

This paper is organized as follows: in Sec. 2 we briefly review different
strong-coupling perturbation theories and the high temperature quantum
virial expansion theory. A comparison for the bulk universal function
$h(\zeta)$ of a homogeneous Fermi gas at unitarity is presented in
Sec. 3. The validity of the quantum virial expansion in the uniform
case is discussed. In the Sec. 4, we explain how to reconstruct the
trapped universal thermodynamic functions from $h(\zeta)$, and present
a systematic comparison of different strong-coupling theories with
the accurate experimental data, for various thermodynamic functions.
We also examine the applicability of the quantum virial expansion
to a trapped Fermi gas at unitarity. In Sec. 5, we calculate the actual
temperature as a function of the non-interacting temperature at the
same entropy. A summary and outlook is given in Sec. 6.

\section{Theoretical review}

\label{II}

In this section we review two types of strongly interacting Fermi
theories: the strong-coupling perturbation theories and a controllable
quantum cluster expansion theory. We consider a two-component Fermi
gas with equal spin populations. At ultracold temperatures ($<100$nK),
the interatomic interactions between atoms with unlike spins can be
well-described by an \textit{s}-wave scattering length $a_{s}$. For
the case of molecule formation with a broad Feshbach resonance, a
simple two-species model that neglects the molecular field is very
accurate, otherwise a full three-species model is necessary \cite{KheruntsDrummond,Holland-Kokkelmans-2001-2002}.
The hamiltonian of the system can then be written as,\begin{equation}
\mathcal{H}=\sum_{\mathbf{k\sigma}}\left(\epsilon_{\mathbf{k}}-\mu\right)c_{\mathbf{k}\sigma}^{+}c_{\mathbf{k}\sigma}+U\sum_{\mathbf{kk'q}}c_{\mathbf{k}+\mathbf{q}\uparrow}^{+}c_{\mathbf{k'}-\mathbf{q}\downarrow}^{+}c_{\mathbf{k'}\downarrow}c_{\mathbf{k}\uparrow},\label{hami}\end{equation}
 where $\epsilon_{{\bf k}}=\hbar^{2}{\bf k}^{2}/(2m)$ is the fermionic
kinetic energy at wave number $k$, and \begin{equation}
\frac{1}{U}=\frac{m}{4\pi\hbar^{2}a_{s}}-\sum_{\mathbf{\mathbf{{\bf k}}}}\frac{1}{2\epsilon_{\mathbf{k}}}\end{equation}
 is the \emph{bare} contact interaction renormalized in terms of the
\textit{s}-wave scattering length $a_{s}$.

\subsection{Strong-coupling perturbation theories}

\label{IIa}

\begin{figure}[htp]

\begin{centering}
\includegraphics[clip,width=0.70\textwidth]{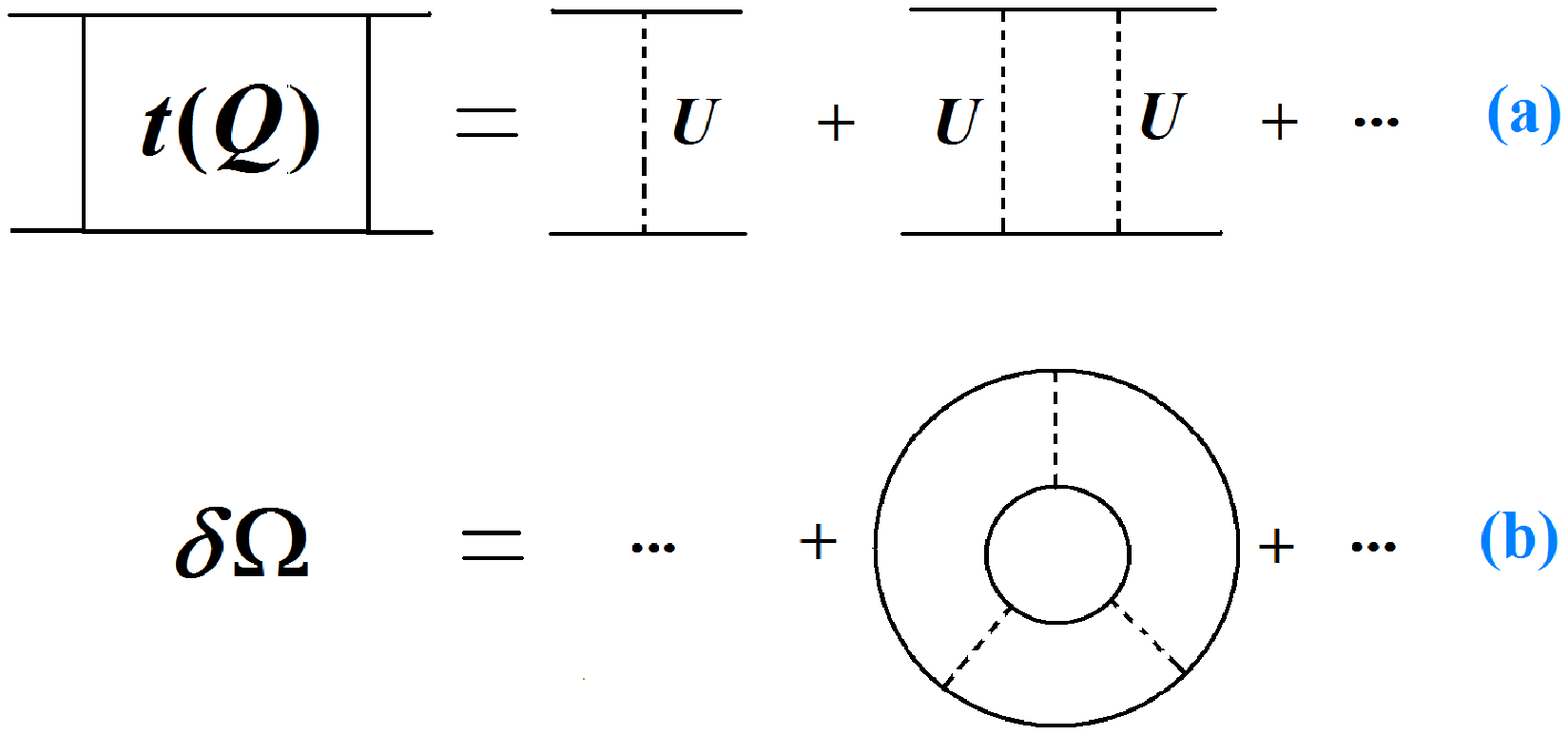} 
\par\end{centering}

\caption{Diagrammatic representation of \textit{T}-matrix and pair fluctuation
contribution to the thermodynamic potential. The solid line represents
the single-particle Green's function, while the dashed line shows
the bare contact interaction. Note that, these diagrams are valid
at all temperatures, both below and above the normal-superfluid transition.
However, below the transition, the single-particle Green's function
should be a $2\times2$ matrix. }

\label{fig1} 
\end{figure}

We start with a brief overview of the most commonly used strong coupling
theories for a Fermi gas at unitarity. These are approximate many-body
\textit{T}-matrix theories \cite{flucttheory,flucttheory2}, since no exact results
are known in three dimensions. Such theories typically go beyond BCS
theory by including an infinite set of higher order Feynman diagrams.
The diagrams included, known as the ladder sum in the particle-particle
channel, are still not the complete set of all possible terms in perturbation
theory. However, it is generally accepted that a ladder sum is necessary
in order to include the strong pair fluctuations in the strongly interacting
regime. This is expected to be the leading class of these infinite
sets of diagrams \cite{flucttheory,flucttheory2}. However, there are differences
in the procedures used to obtain the relevant diagrams that are included.

In more detail, we show in Fig. 1a the diagrammatic structure of the\textit{
T}-matrix \cite{flucttheory,flucttheory2}, $t(Q)$, where the \emph{sucessive}
two-particle scattering between fermions with unlike spins is taken
into account to infinite order. This forms a ladder structure, with
the solid line and dashed line representing, respectively, the single-particle
Green function $G$ and the interaction $U$. Consequently, as an
effective interaction the\textit{ T}-matrix can be diagrammatically
represented by, \begin{equation}
t\left(Q\right)=U+UGGU+UGGUGGU+\cdots,\end{equation}
 by summing all the successive scattering process. In the \emph{normal}
state with contact interactions, the ladder sum can be conveniently
calculated as,

\begin{equation}
t\left(Q\right)=\frac{U}{\left[1+U\chi\left(Q\right)\right]}.\label{tmatrix}\end{equation}
 Here and throughout, $Q=({\bf q},i\nu_{n})$, $K=({\bf k},i\omega_{m})$,
and ${\bf q}$ and ${\bf k}$ are wave vectors, while $\nu_{n}=2n\pi k_{B}T$
and $\omega_{m}=(2n+1)\pi k_{B}T$ ($n=0,\pm1,\pm2,\cdots$) are bosonic
and fermionic Matsubara frequencies, respectively.

Different \textit{T}-matrix theories differ in their choice of the
particle-particle propagator $\chi\left(Q\right)$, \begin{equation}
\chi\left(Q\right)=\sum\nolimits _{K}G_{\alpha}\left(K\right)G_{\beta}\left(Q-K\right),\label{propagator2p}\end{equation}
 and the associated self-energy, \begin{equation}
\Sigma\left(K\right)=\sum\nolimits _{Q}t\left(Q\right)G_{\gamma}\left(Q-K\right),\label{selfenergy}\end{equation}
 where we have introduced an energy-momentum summation, $\sum_{K}=$
$k_{B}T\sum_{\omega_{m}}\sum_{{\bf k}}$. The subscripts $\alpha$,
$\beta$, and $\gamma$ in the above equations may either be set to
{}``0'', indicating a non-interacting Green's function \begin{equation}
G_{0}(K)=\frac{1}{i\omega_{m}-\hbar^{2}{\bf k}^{2}/2m+\mu},\end{equation}
 or be absent, indicating a fully dressed interacting Green's function.
In the latter case the Dyson equation, \begin{equation}
G\left(K\right)=\frac{G_{0}\left(K\right)}{\left[1-G_{0}\left(K\right)\Sigma\left(K\right)\right]},\label{dysoneq}\end{equation}
 is required to self-consistently determine $G$ and $\Sigma$. The
only free parameter, the chemical potential $\mu$, is fixed by the
number equation, $N=2 \lim_{\tau\rightarrow 0^+} \sum\nolimits _{K}G(K)e^{i\omega_m \tau}$ \cite{footnote0}. By taking different
combinations of $\alpha$, $\beta$ and $\gamma$, there are six distinct
choices of the \textit{T}-matrix, for which a notation of $(G_{\alpha}G_{\beta})G_{\gamma}$
will be used. As noted earlier, there is no known \emph{a priori}
theoretical justification for which \textit{T}-matrix approximation
is the most appropriate.

It is important to note that, while having the same diagrammatic structure,
the \textit{T}-matrix above and below the superfluid transition temperature
$T_{c}$ are different, due to the use of different Green's functions
$G_{0}$ or $G$. In the superfluid phase below $T_{c}$, the Green's
function has to be a $2\times2$ matrix, accounting for $U(1)$ symmetry
breaking. Accordingly, an additional parameter, the order parameter,
appears.

The simplest choice, $(G_{0}G_{0})G_{0}$, was pioneered by Nozières
and Schmitt-Rink for a normal interacting Fermi gas \cite{nsr}, with
a truncated Dyson equation for the Green's function, \textit{i.e.},
\begin{equation}
G\left(K\right)=G_{0}\left(K\right)+G_{0}\left(K\right)\Sigma\left(K\right)G_{0}\left(K\right)\,.\end{equation}
 This was shown to be equivalent to including the Gaussian pair fluctuations
in the grand thermodynamic potential \cite{sademelo,engelbrecht},
which is shown diagrammatically in Fig. 1b. The NSR theory was extended
recently to the broken-symmetry superfluid phase by several authors
\cite{ohashi,perali,hld1,footnote1,taylor}. However, some of these
approaches involved additional assumptions to reduce computational
difficulties.

A full extension of the original idea of Nozières and Schmitt-Rink
to the below threshold regime was reported by the present authors
\cite{hld1}, with the use of a mean-field ($2\times2$ matrix) BCS
Green's function as {}``$G_{0}$'' in the thermodynamic potential
in Fig. 1b. In the following we shall refer to this extension as a
Gaussian pair fluctuation or GPF approach. Numerical calculations
were then performed at the BEC-BCS crossover for the equation of state
of a homogeneous Fermi gas. Compared to the zero-temperature QMC simulation
for ground state energy \cite{astrakharchik}, we found that the extended
GPF approach works extremely well in the superfluid phase. It provides
a \emph{quantitatively} reliable description of the low-temperature
thermodynamics of a strongly interacting fermionic superfluid.

In greater detail, in our theory below the critical temperature, the
contribution of \textit{T}-matrix pair fluctuations to the thermodynamical
potential takes the form (see, for example, Fig. 1b), \begin{equation}
\delta\Omega=\frac{1}{2}\sum_{Q}\ln\det\left[\begin{array}{cc}
\chi_{11}\left(Q\right) & \chi_{12}\left(Q\right)\\
\chi_{12}\left(Q\right) & \chi_{11}\left({\bf -}Q\right)\end{array}\right],\end{equation}
 where \begin{eqnarray}
\chi_{11} & = & \frac{m}{4\pi\hbar^{2}a_{s}}+\sum\limits _{K}{\cal G}_{11}(Q-K){\cal G}_{11}(K)-\sum_{{\bf k}}\frac{1}{2\epsilon_{{\bf k}}},\nonumber \\
\chi_{12} & = & \sum\limits _{K}{\cal G}_{12}(Q-K){\cal G}_{12}(K),\end{eqnarray}
 are respectively the diagonal and off-diagonal parts of the pair
propagator. Here, ${\cal G}_{11}$ and ${\cal G}_{12}$, used in Fig.
1b for the single-particle line, are BCS Green's functions with a
variational order parameter $\Delta$. Together with the mean-field
contribution \begin{equation}
\Omega_{0}=\sum_{{\bf k}}\left[\epsilon_{{\bf k}}-\mu+\frac{\Delta^{2}}{2\epsilon_{{\bf k}}}+2k_{B}Tf(-E_{{\bf k}})\right]-\frac{m\Delta^{2}}{4\pi\hbar^{2}a_{s}},\end{equation}
 where the excitation energy $E_{{\bf k}}=[(\epsilon_{{\bf k}}-\mu)^{2}+\Delta^{2}]^{1/2}$
and the Fermi distribution function $f(x)=1/(1+e^{x/k_{B}T})$, we
obtain the full thermodynamic potential $\Omega=\Omega_{0}+\delta\Omega$.

All the thermodynamic functions, including the total energy $E$ and
total entropy $S$, can then be calculated straightforwardly following
thermodynamic relations. For consistency, in our formalism we determine
the order parameter using the gap equation $\partial\Omega_{0}/\partial\Delta=0$.
Together with the number equation, $n=-\partial\Omega/\partial\mu$,
we solve iteratively the two parameters $\mu$ and $\Delta$. This
theory with bare BCS Green functions in the pair propagators constitutes
the simplest universal description of strongly interacting fermions,
including the essential contribution from the low-lying collective
Bogoliubov-Anderson modes. As such, this type of theory may have useful
applications to other types of strongly interacting fermionic superfluids.

The GPF and NSR approximation does not attempt to be self-consistent.
More sophisticated strong-coupling theories can be obtained by using
dressed Green functions as pair propagators. For example, one may
consider a $(GG)G$ approximation, with a fully self-consistent propagator.
This was investigated in detail by Haussmann \textit{et al.} \cite{gg94,gg07},
both above and below the superfluid transition temperature. One advantage
of the self-consistent $(GG)G$ approximation is that the theory satisfies
the so-called $\Phi-$derivable approach to the many-body problem
due to Luttinger and Ward, in which the exact one-particle Green functions
play the role of an infinite set of variational parameters. The $(GG)G$
theory is thus conserving.

An intermediate scheme with an asymmetric form for the particle-particle
propagator, \textit{i.e.}, $(GG_{0})G_{0}$, has been discussed in
a series of papers by Levin and co-workers \cite{chen}, based on
the assumption that the treatment of fluctuations should be consistent
with the simpler BCS theory at low temperatures. Although the $(GG_{0})G_{0}$
theory has been explored numerically to some extent \cite{maly},
a complete numerical solution is difficult. A simplified version was
introduced based on a decomposition of the \textit{T}-matrix $t(Q)$
in terms of a condensate part and a pseudogap part, leading to the
so-called {}``pseudogap'' crossover theory \cite{chen}. In the
present comparative study, we will include both the pseudogap modification
as well as the full $(GG_{0})G_{0}$ theory. Due to numerical difficulties,
we shall consider the full $(GG_{0})G_{0}$ theory in the normal phase
only.

It is clear that in the GPF approximation one omits infinite diagrams
that are responsible for the multiparticle interactions. The fully
self-consistent $(GG)G$ theory and partially self-consistent $(GG_{0})G_{0}$
theory attempt to correct for this, by modifying one or more single-particle
Green's function in the diagrams. However, the more crucial interaction
vertices remain unchanged. For brevity, hereafter we shall refer to
the fully self-consistent $(GG)G$ theory and partially self-consistent
$(GG_{0})G_{0}$ theory as $GG$ and $GG_{0}$ theory, respectively.

To close the subsection, we emphasize again that there are currently
no general grounds to decide which strong-coupling theory is the most
appropriate, due to the absence of a small controllable interaction
parameter. However, as we shall see, these approaches do give distinct
predictions which can be tested experimentally.

\subsection{Scale-invariance and universal relation at unitarity}

\label{IIb}

In the unitarity limit, due to the infinitely large scattering length,
the interatomic distance becomes the \emph{only} relevant length scale
in the problem. The internal energy and entropy of the system therefore
scale like, \begin{eqnarray}
E & = & N\epsilon_{F}f_{E}\left[\frac{T}{T_{F}}\right],\label{energyscale}\\
S & = & Nk_{B}f_{S}\left[\frac{T}{T_{F}}\right],\label{entropyscale}\end{eqnarray}
where $T_{F}=\epsilon_{F}/k_{B}$ is the Fermi temperature, and $f_{E}$
and $f_{S}$ are two dimensionless universal functions. The scaling
form leads to a well-known scaling identity in free space: \begin{equation}
\Omega=-\frac{2}{3}E,\label{scalerelation}\end{equation}
which holds as well for any ideal, non-interacting quantum gases.
To show this, we note that at unitarity the pressure of the gas can
be readily determined from $P=-\left[\partial E/\partial V\right]_{N,S}$.
From the expression of entropy (\ref{entropyscale}), it is easy to
see that holding the entropy invariant is equivalent to fixing the
reduced temperature $T/T_{F}$. Hence, the only dependence of the
energy on volume is through the Fermi energy, i.e., $E\propto V^{-2/3}$.
Taking the derivative with respect to the volume, one finds that $P=2E/(3V)$
or $\Omega=-2E/3$. This simple equation relates the pressure or thermodynamic
potential and energy for a strongly interacting Fermi gas at unitarity
in the same way as for its ideal, noninteracting counterpart, although
the energy would be quite different.

The scaling identity Eq. (\ref{scalerelation}) follows naturally
from the thermodynamic relations. At this point, only the GPF (NSR)
approach and the fully self-consistent $GG$ theory satisfy it, since
in both theories we can write down a well-defined thermodynamic potential
and then derive from it other thermodynamic quantities in a consistent
way. Other strong-coupling theory, more or less, runs into the thermodynamic
inconsistencies, and therefore violates Eq. (\ref{scalerelation}).
We note that in the superfluid phase, an \emph{ad-hoc} renormalization
of the interaction strength is required in the fully self-consistent
$GG$ theory, in order to obtain a gapless phonon spectrum \cite{gg07}.
Hence, the $GG$ theory in the supefluid phase does not appear to
be universal without modification. In the calculations with the pseudogap
theory and $GG_{0}$ theory, we shall obtain the energy and entropy
from the chemical potential by integrating the thermodynamic relation
$n=-\partial\Omega/\partial\mu$, in order to satisfy Eq. (\ref{scalerelation}).
The detailed procedure was outlined in Ref. \cite{hld3}.

In the experimental situation with a harmonic trapping potential $V(x)$,
potential energy enters into the total energy. Thus, the scaling identity
should be modified accordingly: \begin{equation}
\Omega=-\frac{1}{3}E.\label{scalerelationtrap}\end{equation}
This is because the potential energy $N\left\langle V\right\rangle $
equals to the internal energy $(3/2)\int d{\bf x}P\left({\bf x}\right)$
in harmonic traps and then the internal energy is a half of the total
energy. The prefactor of 2/3 in Eq. (\ref{scalerelation}) is therefore
reduced by a factor of 2. 

To prove the equality, we may treat the gas as a collection of many
locally equilibrium uniform cells (\textit{i.e.}, using the local
density approximation). The local force balance arising from the pressure
$P\left({\bf x}\right)$ and the trapping potential $V\left({\bf x}\right)$
gives rise to, \begin{equation}
{\bf \nabla}P\left({\bf x}\right)+n\left({\bf x}\right){\bf \nabla}V\left({\bf x}\right)=0.\end{equation}
Taking a inner product of the above equation with ${\bf x}$ and integrating
over the whole space, we readily obtain $N\left\langle V\right\rangle =(3/2)\int d{\bf x}P\left({\bf x}\right)$,
after the use of ${\bf x\cdot\nabla}V\left({\bf x}\right)=2V\left({\bf x}\right)$
for a harmonic trap and the integration by part for ${\bf x\cdot\nabla}P\left({\bf x}\right)$.
Hence, the strongly interacting Fermi gas at unitary obeys the same
virial theorem as for an ideal quantum gas.

\subsection{High temperature quantum cluster expansion}

\label{IIc}

At high temperatures there is a controllable, small parameter, given
by the fugacity $z=\exp(\mu/k_{B}T)\ll1$. This is small because the
chemical potential $\mu$ diverges to $-\infty$ at large temperatures
$T$. In principle, all thermodynamic properties of a interacting
Fermi gas can be cluster expanded in powers of fugacity \cite{ho3,ourve},
even in the strongly interacting limit.

The thermodynamic potential $\Omega^{(1)}$ of an ideal, non-interacting
uniform Fermi gas takes the form, \begin{equation}
\frac{\Omega^{(1)}}{V}=-\frac{2k_{B}T}{\lambda^{3}}\left[z+b_{2}^{(1)}z^{2}+\cdots+b_{n}^{(1)}z^{n}+\cdots\right],\label{idealth}\end{equation}
 where $\lambda\equiv[2\pi\hbar^{2}/(mk_{B}T)]^{1/2}$ is the thermal
wavelength and $b_{n}^{(1)}=(-1)^{n+1}/n^{5/2}$ is the $n$-th virial
coefficient. We use the superscript {}``$1$'' to indicate the non-interacting
systems. While at the first glance Eq. (\ref{idealth}) may be valid
at $z<1$ only, its applicability is actually much wider.
The expansion is meaningful for arbitrary positive values of fugacity
through an analytic continuation across the point $z=1$. It is then
possible to show that, as expected for a non-interacting Fermi gas,
\begin{equation}
\frac{\Omega^{(1)}}{V}=-\left(2k_{B}T/\lambda^{3}\right)(2/\sqrt{\pi})\int_{0}^{\infty}t^{1/2}\ln\left(1+ze^{-t}\right)dt\,.\end{equation}

In the presence of interactions, the virial coefficients are modified.
The thermodynamic potential can instead be written as, \begin{equation}
\frac{\Omega-\Omega^{(1)}}{V}=-\frac{2k_{B}T}{\lambda^{3}}\left[\Delta b_{2}z^{2}+\cdots+\Delta b_{n}z^{n}+\cdots\right],\label{intth}\end{equation}
 where $\Delta b_{n}=b_{n}-b_{n}^{(1)}$. Our key assumption in using
the quantum cluster expansion is that the expansion of $\Omega-\Omega^{(1)}$
might be applicable near to the critical temperature, despite the
fact that the fugacity may already be much larger than unity close
to the superfluid transition. It is possible to test this conjecture
in either BCS or BEC limit, by analytically calculating the virial
coefficient $\Delta b_{n}$ and hence the radius of convergence of
the expansion. We leave this possibility in a future study.

At unitarity, the virial coefficients are temperature \emph{independent}
and are known up to the fourth order: $\Delta b_{2}=1/\sqrt{2}$,
$\Delta b_{3}=-0.35501298$, and $\Delta b_{4}\simeq0.096\pm0.015$.
The second virial coefficient was already known 70 years ago \cite{ho3}.
The third coefficient was calculated recently by the present authors
\cite{ourve}. This theoretical prediction was confirmed experimentally
in the accurate thermodynamic measurements of Nascimbène \textit{et
al.} \cite{ens}. These recent experiments were also able to determine
the fourth coefficient empirically \cite{ens}. With these coefficients,
using $\zeta=z^{-1}$ the universal function $h(\zeta)$ at high temperatures
may be written as, \begin{equation}
h(\zeta)=1+\frac{\Delta b_{2}\zeta^{-2}+\Delta b_{3}\zeta^{-3}+\Delta b_{4}\zeta^{-4}}{\left(2/\sqrt{\pi}\right)\int_{0}^{\infty}t^{1/2}\ln\left(1+\zeta^{-1}e^{-t}\right)dt}.\label{hzve}\end{equation}

The above discussion for a uniform gas can be easily extended to the
case with a harmonic trap: $V_{trap}({\bf r})=m\omega^{2}r^{2}/2$
or, in the general case of a trap with cylindrical symmetry, \begin{equation}
V_{trap}({\bf r})=m\omega_{\perp}^{2}\rho^{2}/2+\omega_{z}^{2}z^{2}/2\end{equation}
 with $\omega\equiv(\omega_{\perp}^{2}\omega_{z})^{1/3}$. Within
the local density approximation, the thermodynamic potential becomes
position-dependent through a local chemical potential $\mu({\bf r})=\mu_{0}-V_{trap}({\bf r})$
or a local fugacity $z(r)=e^{\beta\mu({\bf r})}$. Using the fact
that the virial coefficients are constant at unitarity, the total
(integrated) thermodynamic potential $\Omega_{trap}(\mu_{0},T)=\int d{\bf r}\Omega({\bf r})$
takes the from, \begin{equation}
\Omega_{trap}=-\frac{2\left(k_{B}T\right)^{4}}{\left(\hbar\omega\right)^{3}}\left[z_{0}+\cdots+\left(\frac{b_{n}}{n^{3/2}}\right)z_{0}^{n}+\cdots\right],\label{trapth}\end{equation}
 where $z_{0}=e^{\beta\mu_{0}}$ is the fugacity at the trap center.
It is easy to see that the $n$th virial coefficient in a trap, $b_{n,trap}=b_{n}/n^{3/2}$,
is much reduced with respect to its uniform counterpart.

The thermodynamic potential of an ideal trapped Fermi gas is now given
by, \begin{equation}
\Omega_{trap}^{(1)}=-\left(k_{B}T\right)^{4}/\left(\hbar\omega\right)^{3}\int_{0}^{\infty}t^{2}\ln\left(1+z_{0}e^{-t}\right)dt\,\,.\end{equation}
 In analogy with the uniform case, we may define a universal function
$h_{trap}(\zeta_{0})=\Omega_{trap}(\mu_{0},T)/\Omega_{trap}^{(1)}(\mu_{0},T)$,
where $\zeta_{0}=e^{-\beta\mu_{0}}$ is the inverse fugacity at the
trap center. Using the expansion at high temperatures, we find explicitly
that,

\begin{equation}
h_{trap}(\zeta_{0})=1+\frac{2\sqrt{2}\Delta b_{2}\zeta_{0}^{2}+8\Delta b_{3}\zeta_{0}/\left(3\sqrt{3}\right)+\Delta b_{4}}{4\zeta_{0}^{4}\int_{0}^{\infty}t^{2}\ln\left(1+\zeta_{0}^{-1}e^{-t}\right)dt}.\label{traphzve}\end{equation}
 
 This expression will be used later on in a comparison for a trapped
Fermi gas at unitarity.

\section{Comparisons for a uniform Fermi gas at unitarity}

\label{III}

We consider now the comparison between theory and experiment for a
uniform Fermi gas at unitarity. For this purpose, we calculate the
universal function $h(\zeta)$ using different perturbation theories
and compare the results to the experimental measurement (Fig. 3a in
Ref. \cite{ens}) This can be easily done with the known equation
of state of the uniform unitarity gas, that is, $\mu\left(T\right)=\epsilon_{F}f_{\mu}(T/T_{F})$
and $E(T)=N\epsilon_{F}f_{E}(T/T_{F})$, where $f_{\mu}$ and $f_{E}$
are two dimensionless functions that depends on the reduced temperature
$\tau=T/T_{F}$ only. Numerically, for a fixed reduced temperature
$\tau$, we calculate the inverse fugacity $\zeta$ and $\Omega^{(1)}\left(\zeta\right)/(N\epsilon_{F})$.
We then obtain, \begin{eqnarray}
\zeta & = & \exp\left[-\frac{f_{\mu}(\tau)}{\tau}\right],\nonumber \\
h(\zeta) & = & \frac{4}{9}\frac{f_{E}(\tau)}{\tau^{5/2}\int_{0}^{\infty}t^{1/2}\ln\left(1+\zeta^{-1}e^{-t}\right)dt}.\end{eqnarray}

\subsection{Perturbation theories}

\label{IIIa}

\begin{figure}[htp]

\begin{centering}
\includegraphics[clip,width=0.70\textwidth]{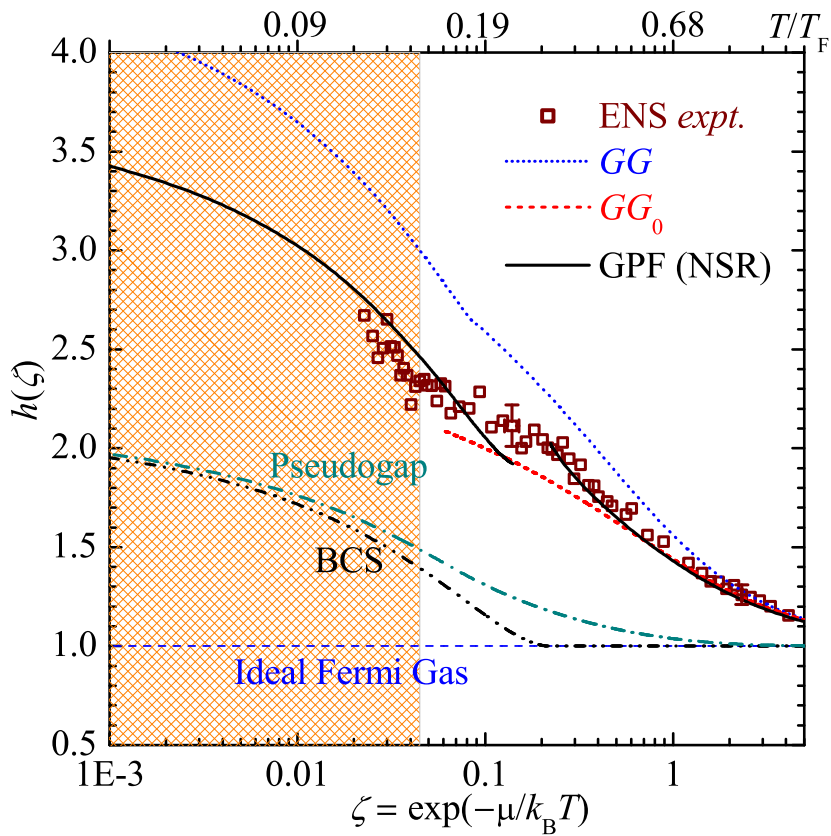} 
\par\end{centering}

\caption{(Color online). Universal thermodynamic function $h(\zeta)$ (squares),
in a uniform gas, compared with predictions from different strong-coupling
theories: the GPF or NSR theory (black solid line), partially consistent
$GG_{0}$ theory (red dashed line), fully self-consistent $GG$ theory
(blue dotted line), pseudogap theory (cyan dash-dotted line), and
the BCS mean-field theory (black dash-dot-dotted line). For an ideal
Fermi gas, $h(\zeta)\equiv1$, as shown by a thin dashed line. The
cross region indicates the superfluid phase. The upper $x$-axis shows
the temperature.}

\label{fig2} 
\end{figure}

The universal functions $h(\zeta)$ obtained in this manner are shown
in Fig. 2 for different perturbation theories, and are compared to
the experimental data (red squares) \cite{ens} without any adjustable
parameters. We indicate the experimentally determined superfluid regime
by thin cross lines, where $\zeta_{c}\simeq0.042$. The $GG$ theory
and the NSR theory above $T_{c}$ were already compared with experiment
by Nascimbène \textit{et al.} in their experimental paper \cite{ens}.
The comparison presented here is much more complete. We find that
the agreement between experiment and the three \textit{T}-matrix perturbation
theories is very good for a large range of temperatures.

In particular, the simplest Gaussian pair fluctuation theory gives
the best quantitative description, though it has a possibly unphysical
discontinuity at the superfluid phase transition. In the Gaussian
pair fluctuation theory, the universal function does not have values
between $[\zeta_{c-}]_{GPF}\simeq0.15$ and $[\zeta_{c+}]_{NSR}\simeq0.22$.
This non-overlap region is mostly caused by the breakdown of the original
NSR theory in the vicinity of transition \cite{hld2}, which predicts
a smaller chemical potential and hence a larger inverse fugacity $\zeta.$
In addtion, the critical inverse fugacity predicted by GPF, $\zeta_{c}\simeq0.2$,
is significantly larger than the experimental observation \cite{ens},
$\zeta_{c}\simeq0.042$. Instead, the self-consistent $GG$ theory
and partially self-consistent $GG_{0}$ theory predict more reasonable
values for $\zeta_{c}$, although their agreement with the experimental
data of $h(\zeta)$ is worse than the GPF (NSR) theory. The prediction of 
available quantum Monte-Carlo simulations is also in close agreement with
the experimental data \cite{bulgac3,burovski2}. Clearly, it
would be useful to have more accurate experimental data at this point,
to better understand the nature of the phase transition.

On the other hand, the pseudogap theory, as a simplification of the
partially self-consistent $GG_{0}$ theory, deviates significantly
from the experimental data. It is therefore not able to capture the
strong fluctuations at unitarity. However, it is certainly better
than the BCS mean-field theory, which completely ignores pairing fluctuations.

As the tempeature decreases to zero ($T\rightarrow0$ and $\zeta\rightarrow0$),
the universal function $h(\zeta\rightarrow0)\rightarrow\xi^{-3/2}$,
where $\xi=1+\beta$ is the universal parameter. Different theories
predict different universal parameters, i.e., $\xi_{BCS}\simeq0.59$,
$\xi_{GPF}\simeq0.401$, and $\xi_{GG}\simeq0.36$. Thus, we find
that, $h_{BCS}(\zeta=0)\simeq2.17$, $h_{GPF}(\zeta=0)\simeq3.94$,
and $h_{GG}(\zeta=0)\simeq4.65$.

\subsection{Virial expansion comparisons}

\label{IIIb}

Let us now focus on the high temperature regime of $\zeta>1$
or $z<1$. A comparison of experimental data to the virial
expansion in Eq. (\ref{hzve}) has already been carried out by Nascimbène
\textit{et al.} \cite{ens}. This led to the confirmation of our theoretical
prediction of the third virial coefficient $\Delta b_{3}\simeq-0.35$
as well as an experimental determination of the fourth virial coefficient
$\Delta b_{4}\simeq0.096\pm0.015$. The third virial calculation requires
an exact solution of a quantum three-body problem, which is known.
However, the exact solution of the quantum four-body problem needed
for the fourth coefficient, is yet to be theoretically obtained.

Here, we show that this accurate experimental data can serve as a
benchmark to determine to what extent the virial expansion is quantitatively
reliable. To be concrete, we shall define the criterion of {}``quantitative''
applicability as an agreement within 10\% relative error for the function
$h(\zeta)-1$, that is, after the non-interacting background is removed
from the universal function $h(\zeta)$. For a {}``qualitative''
applicability, we relax the criterion on the relative error to 50\%.

\begin{figure}[htp]

\begin{centering}
\includegraphics[clip,width=0.70\textwidth]{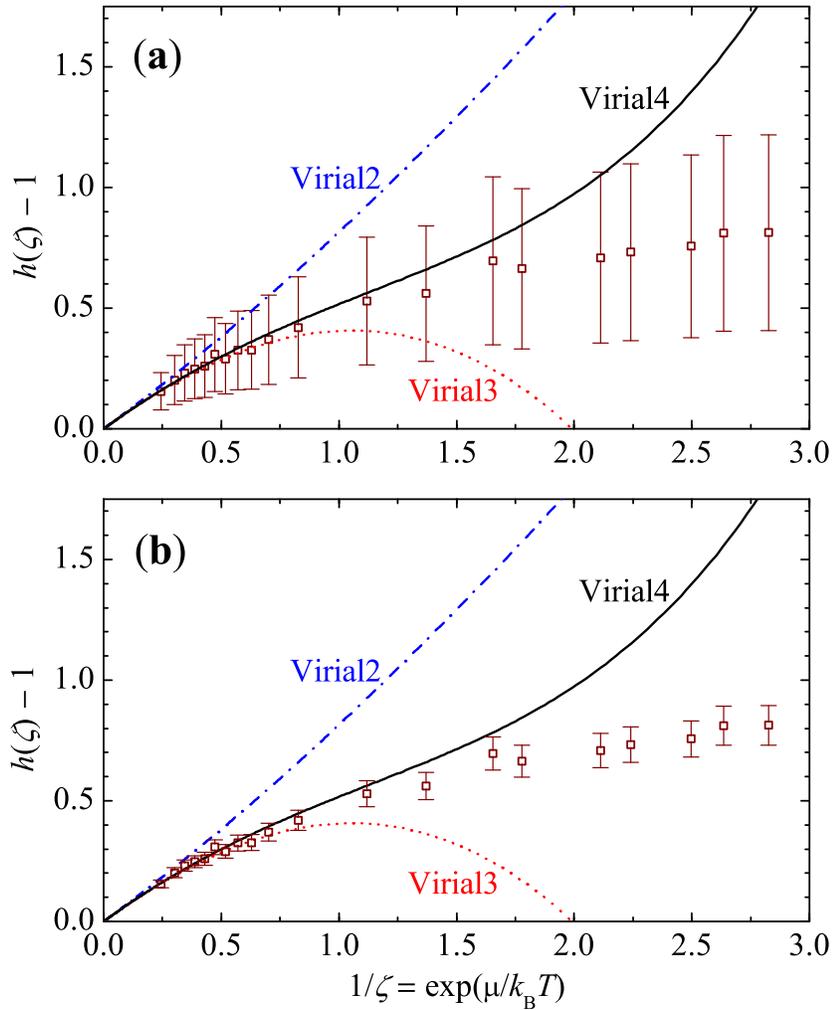} 
\par\end{centering}

\caption{(Color online). Plot of universal function $h(\zeta)-1$ as a function
of fugacity, compared with the experimental data (empty squares).
The error bars in (a) and (b) indicate respectively the 50\% and 10\%
relative errors of $h(\zeta)-1$, in accord with the {}``qualitative''
and {}``quantitative'' criterions as described in the text.}

\label{fig3} 
\end{figure}

Fig. 3 compares the virial expansion predictions (up to the fourth
virial coefficient) for $h(\zeta)-1$ with the experimental data,
using artificial 50\% (a) and 10\% (b) relative errors for comparison
purposes. We are then able to estimate a critical fugacity, below
which the $n$$^{th}$-order virial expansion is either qualitatively
or quantitatively valid. The result is tabulated in Table I, where
the critical fugacities have also been converted to critical temperatures
by using the NSR equation of state, which provides an excellent description
of the experimental data. For the $2$$^{nd}$ virial expansion
that accounts for the leading interaction effect, we determine that
for a homogeneous Fermi gas at unitarity, it is quantitatively and
qualitatively reliable above $T/T_{F}=1.3$ and $T/T_{F}=0.9$, respectively.

\begin{table}[h]
 \begin{tabular}{ccccc}
 &  &  &  & \tabularnewline
\hline
\hline 
Order  & $z_{50}$  & $z_{10}$  & $(T/T_{F})_{50}$  & $(T/T_{F})_{10}$ \tabularnewline
\hline 
Virial2  & 0.7  & 0.4  & 0.87  & 1.30 \tabularnewline
Virial3  & 1.5  & 0.7  & 0.52  & 0.87 \tabularnewline
Virial4  & 2.2  & 1.7  & 0.40  & 0.48 \tabularnewline
\hline
\end{tabular}

\caption{Qualitative ($50\%$) or quantitatively ($10\%$) ranges of reliability
for different order virial expansions in free space, as indicated
by the subscript.}

\end{table}

To close this section, we emphasize that, though the experimental
data for the universal function $h(\zeta)$ are very accurate, without
any prior knowledge of the bulk equation of state we still can not
determine the temperature from the \emph{discrete} data. However,
for a trapped Fermi gas at unitarity, using the bulk $h(\zeta)$ we
can indeed determine all the universal thermodynamic functions such
as $E(T)$ and $S(T)$. This is discussed in detail in the next section.

\section{Comparisons for a trapped Fermi gas at unitarity}

\label{IV}

Let us now turn to the experimental determination of the equation
of state of a Fermi gas in a harmonic trap at unitarity, and a comparison
of this data with theory. The basic idea has already been outlined
by Nascimbène \textit{et al.} in the Supplementary Discussion part
of their paper \cite{ens}.

\subsection{Local density approximation}

\label{IVa}

Consider, for example, the total number of atoms, $N=\int d{\bf r}n({\bf r})$.
Within the local density approximation, we may consider the trap to
be isotropic trap with a trapping frequency $\omega\equiv(\omega_{\perp}^{2}\omega_{z})^{1/3}$,
without loss of generality. Using $n({\bf r})=\partial P\left({\bf r}\right)/\partial\mu\left({\bf r}\right)$
and $\partial\mu\left({\bf r}\right)/\partial r=-m\omega^{2}r$, we
find that, 
\begin{equation} N =\int_{0}^{\infty}dr4\pi r^{2}[\partial P\left({\bf r}\right)/\partial\mu\left({\bf r}\right)] =-4\pi/(m\omega^{2})\int_{0}^{\infty}rdP(r).\end{equation}
 Noting that $P=P^{(1)}h(\zeta)$ , and integrating by parts, we obtain:
\begin{equation}
N=\frac{4\pi}{m\omega^{2}}\int_{0}^{\infty}drP^{(1)}\left[\zeta\left(r\right)\right]h\left[\zeta\left(r\right)\right].\end{equation}
 Using $\zeta\left(r\right)=e^{-\beta\mu(r)}=\zeta_{0}\exp[m\omega^{2}r^{2}/(2k_{B}T)]$,
the integration over radius $r$ can be converted to an integration
over the inverse fugacity. One sees that, \begin{equation}
N=\frac{4}{\sqrt{\pi}}\left(\frac{k_{B}T}{\hbar\omega}\right)^{3}\int_{\zeta_{0}}^{\infty}d\zeta\frac{d\sqrt{\ln\left(\zeta/\zeta_{0}\right)}}{d\zeta}f\left(\zeta\right)h\left(\zeta\right),\end{equation}
 where $f\left(\zeta\right)\equiv(2/\sqrt{\pi})\int_{0}^{\infty}t^{1/2}\ln\left(1+\zeta^{-1}e^{-t}\right)dt$.
Recalling that the Fermi energy of a zero-temperature trapped ideal
Fermi gas is $E_{F}=k_{B}T_{F}=(3N)^{1/3}\hbar\omega$, we may rewrite
the above equation in dimensionless form, \begin{equation}
\frac{T_{F}^{3}}{T^{3}}=\frac{12}{\sqrt{\pi}}\int_{\zeta_{0}}^{\infty}d\zeta\frac{d\sqrt{\ln\left(\zeta/\zeta_{0}\right)}}{d\zeta}f\left(\zeta\right)h\left(\zeta\right).\label{number}\end{equation}
 The total energy of the system may be conveniently calculated by
using the scaling relation, Eq. (\ref{scalerelationtrap}). Thus,
the total energy in a trap is given by \begin{equation}
E  =-3\Omega=12\pi\int_{0}^{\infty}drr^{2}P(r)=12\pi\int_{0}^{\infty}drr^{2}P^{(1)}\left[\zeta\left(r\right)\right]h\left[\zeta\left(r\right)\right]\,\,.\end{equation}
 Converting to the variable $\zeta$, we find that: \begin{equation}
\frac{E}{NE_{F}}=\frac{72}{\pi^{1/2}}\frac{T^{4}}{T_{F}^{4}}\int_{\zeta_{0}}^{\infty}d\zeta\frac{d\sqrt{\ln\left(\zeta/\zeta_{0}\right)}}{d\zeta}\ln\frac{\zeta}{\zeta_{0}}f\left(\zeta\right)h\left(\zeta\right).\label{energy}\end{equation}
 The entropy follows directly from the thermodynamic relation $S=(E-\Omega-\mu_{0}N)/T$.
Using the fact that $\mu_{0}/E_{F}=-(T/T_{F})\ln\zeta_{0}$, we obtain
straightforwardly \begin{equation}
\frac{S}{Nk_{B}}=\frac{4}{3}\frac{T_{F}}{T}\frac{E}{NE_{F}}+\ln\zeta_{0}.\label{entropy}\end{equation}

The coupled equations (\ref{number}), (\ref{energy}) and (\ref{entropy})
determine, respectively, the temperature, energy, and entropy of a
trapped Fermi gas at unitarity as a function of the inverse fugacity
$\zeta_{0}$. In the calculations, we discretize the integral over
$\zeta$ and take the values of $h\left(\zeta\right)$ solely from
the experimental measured data. In this way \cite{ens}, we avoid
the use of any interpolating or fitting function to the experimental
data $h\left(\zeta\right)$. In addition, the statistical error of
the experimental data of $h\left(\zeta\right)$ is reduced. The detailed
procedure for these numerical calculations is given in Appendix A.
For convenience, we shall refer to this trapped equation of state,
re-constructed from $h\left(\zeta\right)$, as the {}``experimental''
measurement or data for the equations of state of a trapped Fermi
gas at unitarity.

It is readily seen that we can calculate the theoretical prediction
for the trapped equation of state by using exactly the same local
density approximation procedure, combined with a theoretical universal
function $h(\zeta)$ generated from different strong-coupling theories
for a uniform Fermi gas. We note that our numerical procedure of calculating
the trapped equation of state is an average procedure integrating
over the trap and is quite insensitive to the smoothness of $h(\zeta)$.
Therefore, even though there is a discontinuity in the theoretical
universal function, as in our Gaussian pair fluctuation theory, we
obtain a much smoother trapped equation of state. As can be seen from
Appendix A, in that case, we simply join linearly between $\left[\zeta_{c-}\right]_{GPF}$
and $\left[\zeta_{c+}\right]_{NSR}$ to remove the discontinuity of
the universal function in the non-overlap region $\left(\left[\zeta_{c-}\right]_{GPF},\left[\zeta_{c+}\right]_{NSR}\right)$.

\subsection{Trapped universal thermodynamics: $E(S)$}

\label{IVb}

In previous work \cite{natphys}, we gave experimental evidence that
any strongly interacting Fermi gases at unitarity has universal thermodynamics.
The energy and entropy relation $E(S)$ measured on $^{6}$Li and
$^{40}$K atomic clouds in three different trapping potentials all
fall precisely on a single curve. The trapped equation of state $E(S)$
deduced from the experimental data of $h(\zeta)$ by Nascimbène \textit{et
al.} \cite{ens} provides an independent check of universality, with
a much improved accuracy, in a fourth different set of experimental
conditions. This is illustrated in Fig. 4, where we plot the new measurements
using green circles. All the four sets of experimental data follow
the theoretical prediction given by the simplest GPF approximation.
In particular, the difference between our theory and the new measurement
is nearly indistinguishable, as shown clearly in Fig. 4b for the low
temperature regime. This gives so far the strongest evidence for fermionic
universality.

\begin{figure}[htp]

\begin{centering}
\includegraphics[clip,width=0.70\textwidth]{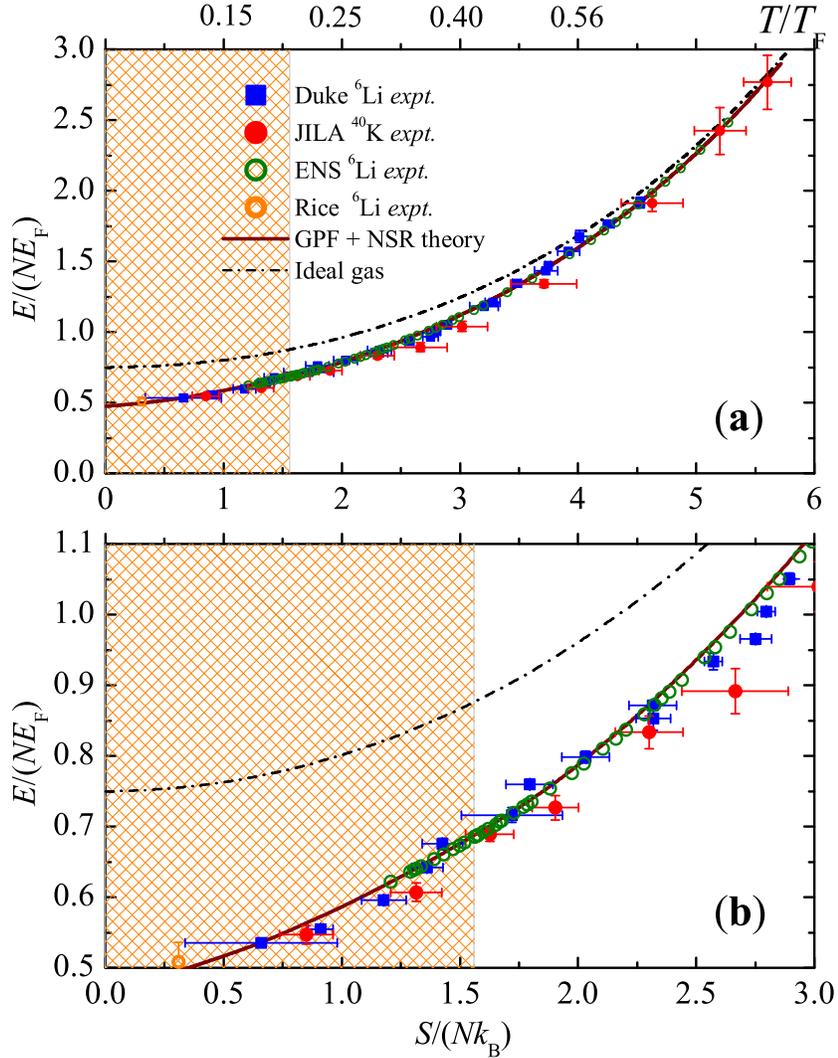} 
\par\end{centering}

\caption{(Color online). $E(S)$\ of a strongly interacting atomic Fermi gas
of either $^{6}$Li or $^{40}$K atoms in different trapping potentials
scale into a single theoretical curve, which is predicted by the GPF
and NSR theory. Lower panel highlights the low temperature regime.
The cross region indicates the superfluid phase below an experimental
critical entropy $(S/Nk_{B})_{c}\simeq1.56$. The determination of
the critical entropy is described in Sec. V. The upper $x$-axis in
(a) plots the temperature.}

\label{fig4} 
\end{figure}

\begin{figure}[htp]

\begin{centering}
\includegraphics[clip,width=0.70\textwidth]{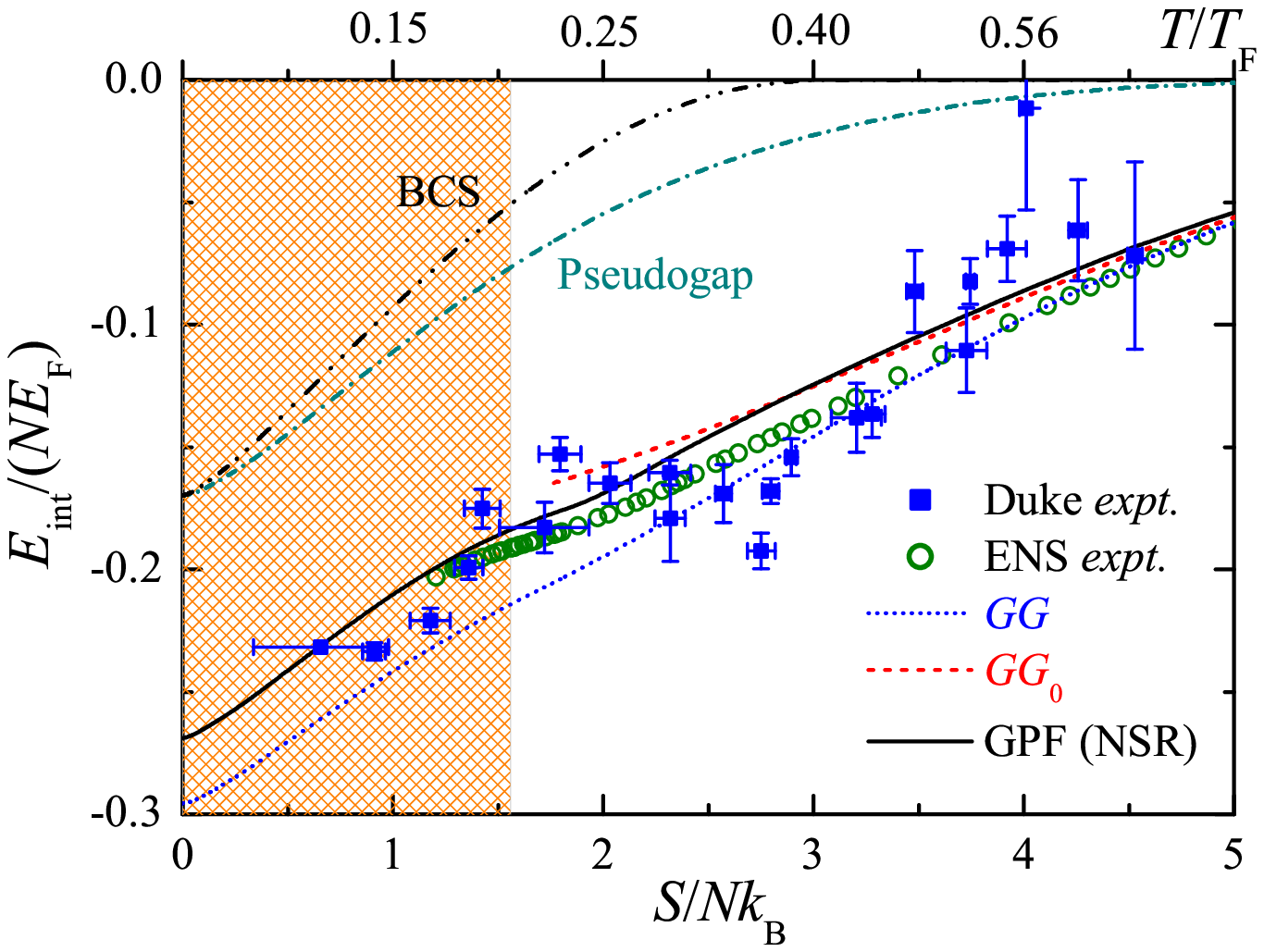} 
\par\end{centering}

\caption{(Color online). Interaction energy as a function of entropy for a
strongly interacting Fermi gas in a harmonic trap. The experimental
data (symbols) are compared with the predictions of three different
strong-coupling theories. The cross region indicates the superfluid
phase below the critical entropy $(S/Nk_{B})_{c}\simeq1.56$. The
temperature at different entropy is plotted on the upper $x$-axis.}

\label{fig5} 
\end{figure}

To better visualize the difference between theory and experiment,
we follow the strategy used in our previous comparative study \cite{hld3}
and calculate the interaction energy $E_{int}=E-E_{IG}$,
which is the difference of energies between an interacting Fermi gas
($E$) and an ideal Fermi gas ($E_{IG}$) at the \emph{same}
entropy. Fig. 5 shows the interaction energy versus entropy in a harmonic
trap as predicted by different strong-coupling theories in comparison
with the experimental data reported at Duke \cite{duke3} and ENS
\cite{ens}. It is impressive that at this much reduced scale, the
new experimental data obtained by Nascimbène \textit{et al.} at ENS
(green empty circles) still appear to be very smooth, suggesting an
absolute statistical error of about $0.01NE_{F}$ or a relative error
of one percent in energy, although there may be systematic errors
at this level.

This error bar is already much smaller than the difference among different
\textit{T}-matrix approximations of GPF, $GG_{0}$ and $GG$, which
is roughly of the order $0.05NE_{F}$. It is clear that below threshold
the GPF approach provides the closest prediction to the new accurate
measurement below threshold, with a difference at most $0.01NE_{F}$.
However, well above threshold the fully self-consistent $GG$ theory
gives better agreement. We note here that the above threshold $GG$
theory is also universal, without \emph{ad-hoc} renormalizations.

\subsection{Trapped universal function $h_{trap}(\zeta_{0})$}

\label{IVc}

In analogy with the comparison between theory and experiment for the
universal function $h(\zeta)$ for a homogeneous Fermi gas at unitarity,
we consider now the comparison for the trapped universal function
$h_{trap}(\zeta_{0})=\Omega_{trap}/\Omega_{trap}^{(1)}$.
Because of the proportionality between energy and thermodynamic potential
at unitarity (i.e., the scaling relation), it is convenient to calculate
$h_{trap}(\zeta_{0})$ using the expression, \begin{equation}
h_{trap}(\zeta_{0})=\frac{E/(NE_{F})}{9\left(T/T_{F}\right)^{4}\int_{0}^{\infty}t^{2}\ln\left(1+\zeta_{0}^{-1}e^{-t}\right)dt},\end{equation}
 where the denominator is simply $E^{(1)}/(NE_{F})$ at given inverse
fugacity $\zeta_{0}$ and temperature $T$.

\begin{figure}[htp]
\begin{centering}
\includegraphics[clip,width=0.70\textwidth]{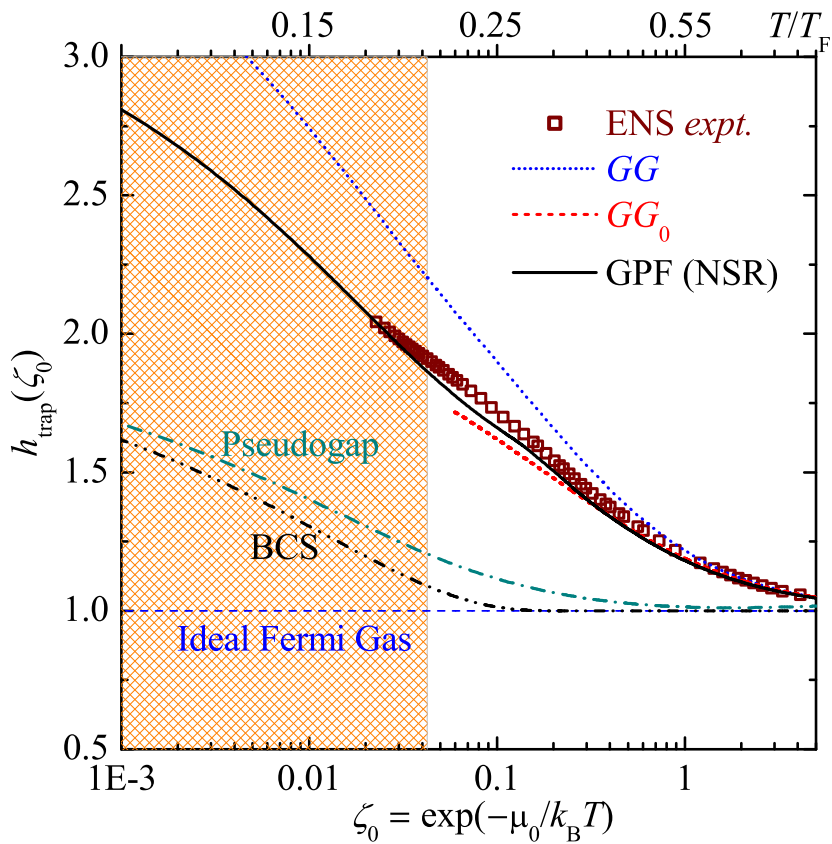} 
\par\end{centering}

\caption{(Color online). Universal function $h_{trap}(\zeta_{0})$ in
a harmonic trap. Here, $\zeta_{0}$ is the inverse fugacity at the
center of the trap. Different theoretical predictions (lines as indicated)
are compared with the experimental measurement (squares). The cross
region indicates the superfluid phase below a critical inverse fugacity
$\left(\zeta_{0}\right)_{c}\simeq0.042$. The upper $x$-axis shows
the temperature.}

\label{fig6} 
\end{figure}

Fig. 6 compares the theoretical predictions for the trapped universal
function from different strong-coupling theories, compared with the
experimental measurement at ENS (empty squares) \cite{ens}. Both
the experimental data and the theoretical GPF (NSR) predictions for
the trapped universal function $h_{trap}(\zeta_{0})$ are now much
smoother after integrating over the trap. Thus, the discontinuity
of the normal-superfluid transition in the GPF prediction of $h(\zeta)$
disappears completely in $h_{trap}(\zeta_{0})$ , due to trap-averaging.
Comparing the different \textit{T}-matrix fluctuation theories in
a trap, one sees that the fully self-consistent $GG$ theory gives
a slightly better agreement with the experimental results than the
GPF and $GG_{0}$ theories at high temperatures, with an inverse fugacity
$\zeta_{0}>2$. At low temperatures where $\zeta_{0}\ll1$, however,
the simplest GPF approach provides the best description.

We note that, in harmonic traps the trapped universal function
$h_{trap}(\zeta_{0}=0)=\xi^{-3/2}$ is exactly the same as
the uniform universal function at zero temperature. Therefore, $h_{trap,BCS}(\zeta_{0}=0)\simeq2.17$,
$h_{trap,GPF}(\zeta_{0}=0)\simeq3.94$, and $h_{trap,GG}(\zeta_{0}=0)\simeq4.65$.
However, these limiting values seem to be difficult to reach, compared
to the uniform case.

\subsection{Quantum virial comparisons: $h(\zeta)$}

\label{IVd}

We now examine the applicability of the quantum virial expansion for
a trapped Fermi gas at unitarity, by using Eq. (\ref{traphzve}) for
the trapped universal function. We use the same idea as in Fig. 3
and the same criterion for {}``qualitative'' and {}``quantitative''
reliability. In Fig. 7, we report the successive virial expansion
as a function of fugacity up to the $4^{th}$ order, compared
with the experimental data. The estimates of the critical fugacity
and of the critical temperature for different orders are tabulated
in Table II. To the leading second order, we find that the expansion
is quantitatively reliable for temperatures down to $T\simeq0.7T_{F}$,
much smaller than we found for a homogeneous Fermi gas at unitarity.

\begin{figure}[htp]

\begin{centering}
\includegraphics[clip,width=0.70\textwidth]{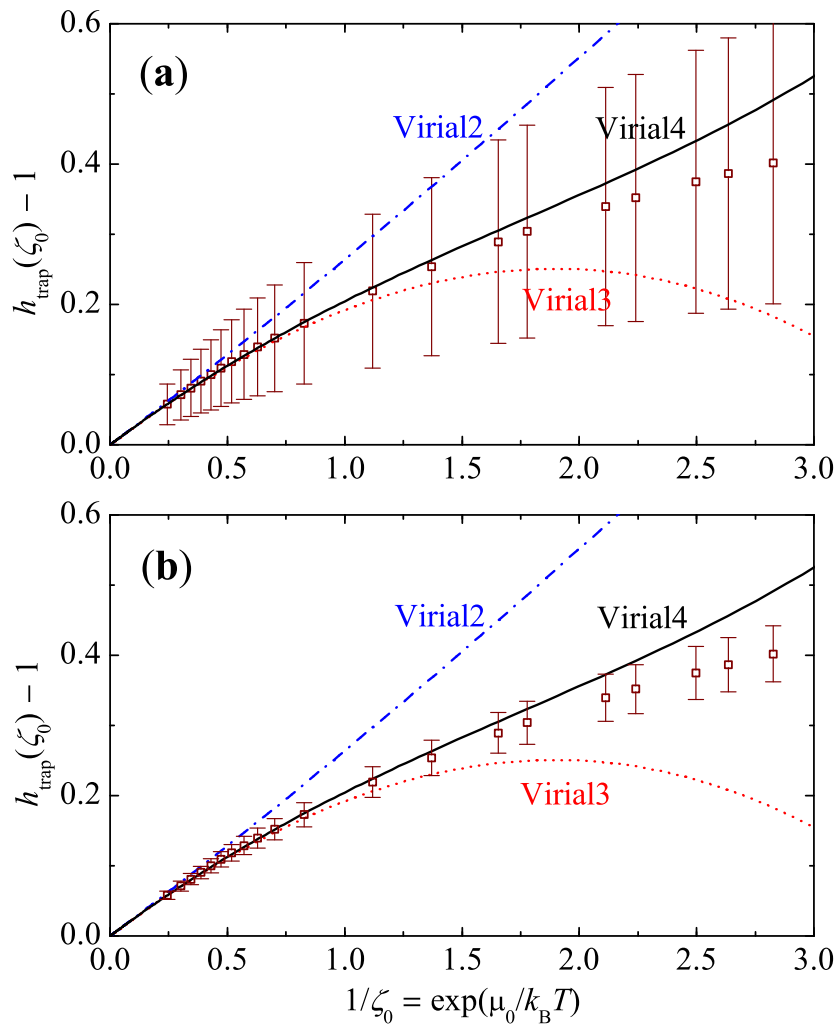} 
\par\end{centering}

\caption{(Color online). Examination of the reliability of the quantum virial
expansion for a trapped Fermi gas at unitarity. This is basically
the same as what shown in Fig.3, but now the plotted curves are computed
for a trapped Fermi gas.}

\label{fig7} 
\end{figure}

\begin{table}[h]
 \begin{tabular}{ccccc}
 &  &  &  & \tabularnewline
\hline
\hline 
Order  & $z_{50}$  & $z_{10}$  & $(T/T_{F})_{50}$  & $(T/T_{F})_{10}$ \tabularnewline
\hline 
Virial2  & 1.5  & 0.5  & 0.45  & 0.66 \tabularnewline
Virial3  & 2.7  & 1.4  & 0.36  & 0.46 \tabularnewline
Virial4  & 3.7  & 2.1  & 0.33  & 0.39 \tabularnewline
\hline
\end{tabular}

\caption{Qualitative ($50\%$) or quantitatively ($10\%$) ranges of reliability
for different order virial expansions in trapping potentials, as indicated
by the subscript.}

\end{table}

This much wider applicability is due to the significantly reduced
higher order virial coefficients in a harmonic trap, i.e., $b_{2,trap}=b_{2}/(2\sqrt{2})$.
It is readily seen that, with inclusion of higher order virial coefficients,
the accuracy of virial expansion can be improved. Up to the known
fourth virial coefficient, we find that the bound for quantitative
applicability decreases further to $T\simeq0.4T_{F}$, which is a
typical experimental temperature for a Fermi gas in its normal state.
We thus show that the quantum virial expansion method is a very useful
tool for understanding the properties of a normal, strongly interacting
Fermi gas in a harmonic trap.

\subsection{Thermodynamic functions $E(T)$ and $S(T)$}

\label{IVe}

As we mentioned earlier, in addition to the energy-entropy relation
$E(S)$, the measurement by Nascimbène \textit{et al. }\cite{ens}\textit{\ }was
able to re-construct a complete set of thermodynamic functions in
harmonic traps, such as $E(T)$ and $S(T)$. This provides us with
a unique opportunity for a systematic comparison between theory and
experiment for a trapped Fermi gas at unitarity, without the use of
any fitting functions or adjustable parameters.

\begin{figure}[htp]
\begin{centering}
\includegraphics[clip,width=0.70\textwidth]{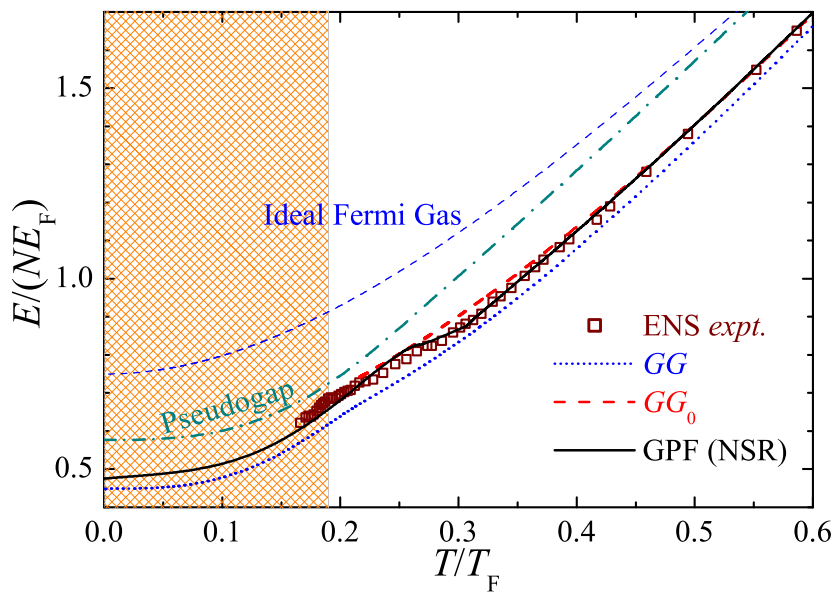} 
\par\end{centering}

\caption{(Color online). Universal thermodynamic function $E(T)$. Different
theoretical preditions (as indicated) are compared with the experimental
data (empty squares). The cross region indicates the superfluid phase
below a criticla temperature $(T/T_{F})_{c}\simeq0.19$. This experimental
critical temperature of a trapped Fermi gas at unitarity is determined
in Sec. V.}

\label{fig8} 
\end{figure}

Fig. 8 presents the comparison for the total energy as a function
of temperature. As anticipated, the simplest GPF approach provides
the best quantitative agreement with the experimental data. However,
the GPF theory predicts a larger normal-superfluid transtion temperature,
$(T/T_{F})_{c}\simeq0.27$, as indicated by a small bump. In contrast,
the less accurate self-consistent $GG$ theory and partially self-consistent
$GG_{0}$ theory predict that $(T/T_{F})_{c}\simeq0.21$, which is
much closer to the experimental observation of $(T/T_{F})_{c}\simeq0.19$.

It is interesting to note that the GPF curve in the figure was first
calculated by the present authors \cite{hld2,footnote2} and was compared
to the heat capacity measurement reported by Kinast \textit{et al.
}\cite{duke2}. However, at that time, the temperature was not independently
calibrated, due to the absence of a reliable thermometry in the strongly
interacting regime. An empirical temperature was used, obtained by
fitting the integrated one-dimensional density profile to an ideal
Thomas-Fermi distribution.

In this earlier comparison, empirical temperatures were converted
to actual temperatures using the pseudogap theory \cite{duke2}. As
a consequence, the resulting experimental data of $E(T)$ appeared
to agree well with the pseudogap theory \cite{duke2}. This is in
sharp contrast to what is shown in Fig. 8, where the pseudogap theory
clearly fails to account for the strong pairing fluctuations at either
low temperatures ($T<0.1T_{F}$) or high temperatures ($T>0.3T_{F}$).
We therefore conclude that while the empirical temperature approach
provides a rough thermometry, its model-dependence and insensitivity
to the actual temperature makes it less useful as a tool for accurate
comparison of theory with experiment.

\begin{figure}[htp]
\begin{centering}
\includegraphics[clip,width=0.70\textwidth]{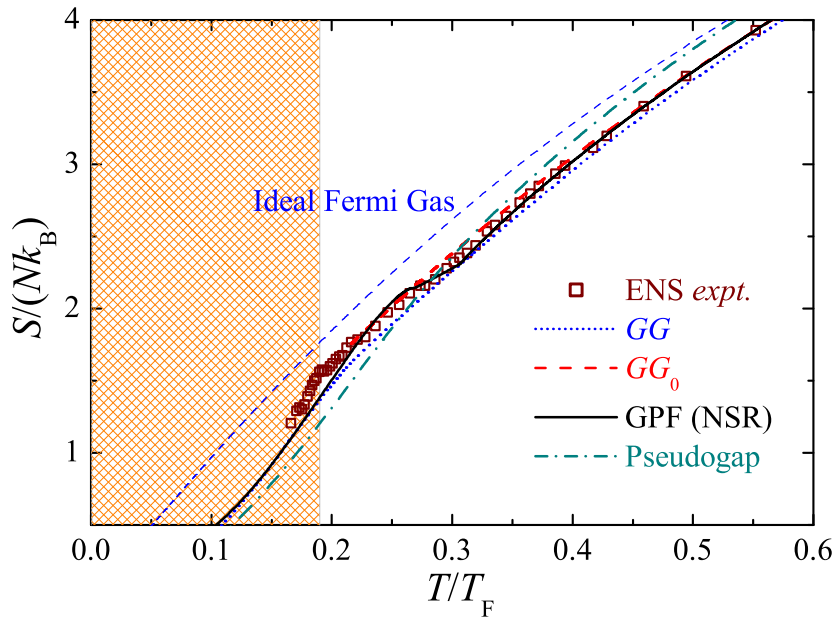} 
\par\end{centering}

\caption{(Color online). Universal thermodynamic function$S(T)$. Different
theoretical preditions (as indicated) are compared with the experimental
data (empty squares). The cross region indicates the superfluid phase.}

\label{fig9} 
\end{figure}

Fig. 9 shows the comparison between theory and experiment for the
entropy as a function of temperature. We find again that the GPF approach
gives an overall best fit to the experimental data. Compared to the
case of total energy, however, the effect of pairing fluctuations
on the entropy is less significant. As a result, all the perturbation
theories predict a similar entropy curve. Their difference to the
ideal Fermi gas prediction (thin dashed line) is also small. This
provides a justification for a recent calibration strategy used for
determining the entropy of a weakly interacting Fermi gas \cite{duke3},
in which the entropy of a $^{6}$Li cloud at a magnetic field $B=1200G$
is assumed to be close to that of an ideal Fermi gas.

\subsection{Quantum virial expansion for $E(T)$ and $S(T)$}

\label{IVe}

\begin{figure}[htp]
\begin{centering}
\includegraphics[clip,width=0.70\textwidth]{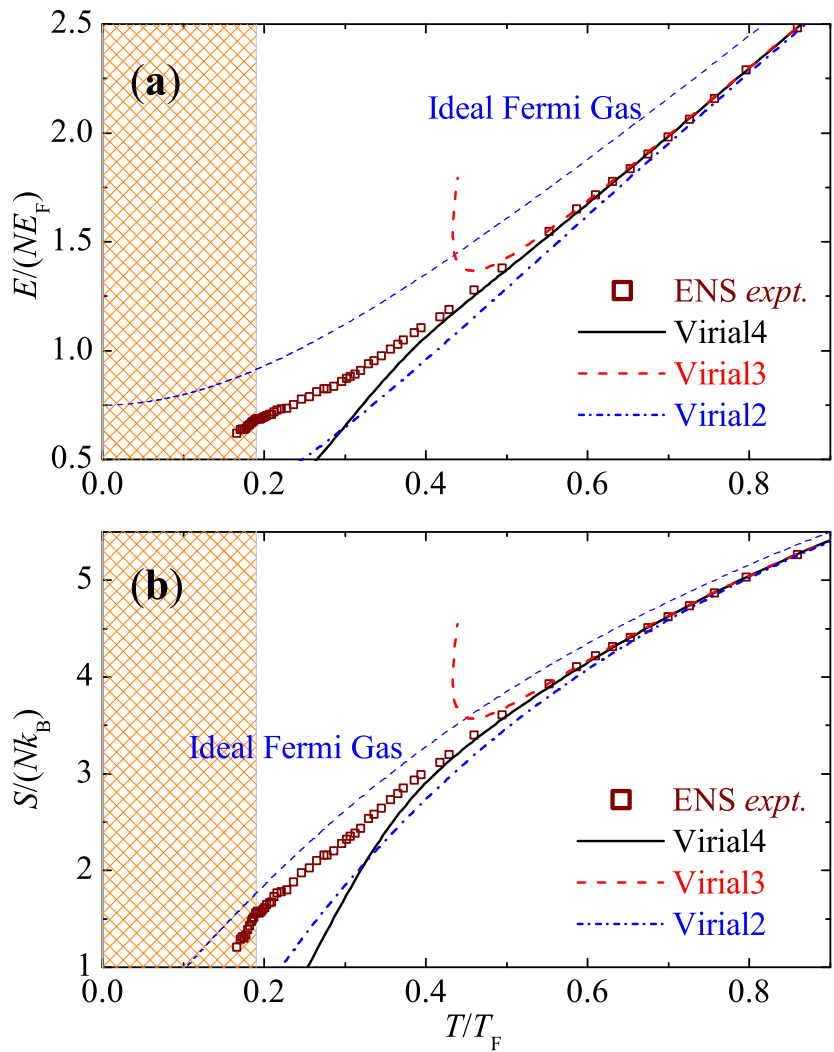} 
\par\end{centering}

\caption{(Color online). $E(T)$ and $S(T)$ for a trapped Fermi gas compared
to the quantum virial expansion predictions. The cross region indicates
the superfluid phase.}

\label{fig10} 
\end{figure}

Finally, we may check the applicability of quantum virial expansion
by using the thermodynamic functions $E(T)$ and $S(T)$. This is
illustrated in Fig. 10, where the theoretical predictions of a virial
expansion up to $4^{th}$ order for $E(T)$ and $S(T)$ are
compared with the experimental data. We may determine directly from
the figures the critical temperature related to the reliability of
virial expansion. These are in agreement with the values listed in
Table II.

\section{Thermometry of a trapped Fermi gas at unitarity}

\label{V}

We have noted that the temperature of a strongly interacting Fermi
gas is difficult to measure experimentally in ultra-cold atom experiments.
Unlike the situation with cryogenic experiments in the past, ultra-cold
atoms are completely insulated by a high vacuum from any external
reservoir at a known temperature. A useful way to quantify the temperature
is to measure a non-interacting temperature $T_{i}$ of an ideal,
non-interacting Fermi gas. This can be easily measured from the density
profile, before a slow, adiabatic sweep to the Feshbach resonance.
Since the entropy of a non-interacting Fermi gas is known, and is
unchanged in an adiabatic sweep, this is essentially an entropy measurement.
This procedure was first adopted by Regal \textit{et al.} at JILA
\cite{jila2}.

\begin{figure}[htp]

\begin{centering}
\includegraphics[clip,width=0.70\textwidth]{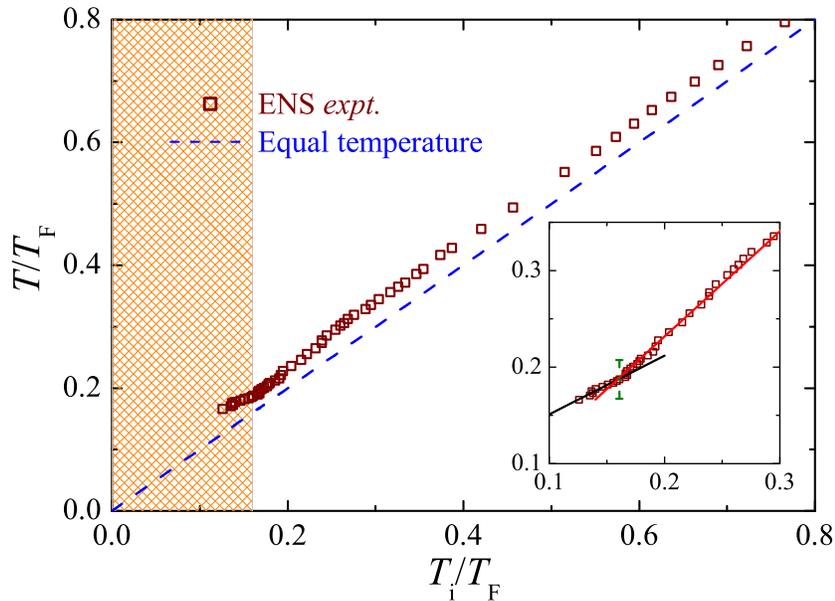} 
\par\end{centering}

\caption{(Color online). $T$ is shown as a function of a non-interacting temperature
$T_{i}$, which can be measured before the adiabatic sweep to the
Feshbach resonance. The blue dashed line denotes equal temperatures.
Inset shows clearly a kink at low temperature regime in an enlarged
scale. By linear-fitting separately the data around the kink, we determine
a characteristic temperature $(T/T_{F})_{0}\simeq0.19\pm0.02$ or $(T_{i}/T_{F})_{0}\simeq0.16\pm0.02$.
The cross region indicates the experimentally determined superfluid phase.}

\label{fig11} 
\end{figure}

The accurate determination of the universal thermodynamic function
$S(T)$ in the last section presents a model-independent way to re-calibrate
the $T_{i}$ thermometry for a trapped Fermi gas at unitarity. By
equating $S(T)$ and $S_{IG}(T_{i})$, where $S_{IG}$
is the entropy of an ideal Fermi gas, we can express the temperature
$T$ of strongly interacting Fermi gases as a function of the non-interacting
temperature $T_{i}$. The result is reported in Fig. 11. The inset
emphasizes the normal-superfluid transition regime. Note that the
isentropic conversion to resonance tends to decrease the temperature
so that $T_{i}$ is always somewhat below the temperature $T$ at
unitarity. This can be seen from the dashed line for which $T=T_{i}$.

We may identify three different temperature regimes from the figure.
At temperatures $T_{i}>0.3T_{F}$, the calibration curve is nearly
parallel with the equal temperature line. To a good approximation,
we find that $T\simeq T_{i}+0.04T_{F}$. Below $T_{i}=0.3T_{F}$,
the unitarity temperature seems to decrease slightly faster with decreasing
non-interacting temperature. However, at a characteristic temperature $(T_{i}/T_{F})_{c}\simeq0.16\pm0.02$,
this trend changes suddenly and we observe that the unitarity temperature
now tends to saturate with any further decrease of the non-interacting
temperature. This interesting feature is clearly seen in the inset,
where we linearly fit the data above and below $(T_{i}/T_{F})_{c}$.

We can identify the sudden change as the deviations of thermodynamic properties
away from normal Landau-Fermi-liquid behavior \cite{ens,bulgac3} and therefore determine a characteristic 
temperature $(T/T_{F})_{0}\simeq0.19\pm0.02$. This value agrees very well with
the one that deduced from the homogeneous critical temperature by
Nascimbène \textit{et al.} \cite{ens} and the condensate fraction measurement by Horikoshi \textit{et al.}  \cite{tokyo}. 
The characteristic non-interacting temperature $(T_{i}/T_{F})_{0}\simeq0.16$ is also 
in very good agreement with the measurement at JILA \cite{jila2} and the recent moment of
inertia measurement at Innsbruck \cite{innsbruck}. Using the experimental
thermodynamic functions $E(T)$ and $S(T)$ in Figs. (8) and (9),
we obtain $(E/NE_{F})_{0}\simeq0.68$ and $(S/Nk_{B})_{0}\simeq1.56$.

\section{Conclusions and outlooks}

\label{VI}

In conclusion, from the experimental data measuring a uniform universal
function $h(\zeta)$ we have deduced a complete set of universal thermodynamic
functions for a trapped Fermi gas at unitarity. These accurate experimental
results provide a unique opportunity to test quantum many-body theories
of strongly interacting Fermi gases. We have presented such a study
by systematically comparing the theoretical predictions from typical
strong-coupling theories with the experimental data. The comparison
has no fitting functions or adjustable parameters. All the model approximations
seem to be fluctuating around and not converging towards the accurate
experimental data. We have found that the simple Gaussian pair fluctuation
theory pioneered by Nozières and Schmitt-Rink \cite{nsr,hld1} provides
the best quantitative description for the universal thermodynamic
properties of energy and entropy. Our comparison also includes a quantum
virial expansion theory (or quantum cluster expansion theory) \cite{ho3,ourve}.
We have investigated in detail the applicability of the expansion
in the quantum degenerate regime.

The experimental universal thermodynamic functions calculated in this
work are extremely useful. For instance, the temperature dependence
of entropy $S(T)$ can be used to calibrate accurately the endpoint
temperatures obtained from an adiabatic sweep of the magnetic field
between the ideal and strongly interacting regimes. Therefore, by
measuring an ideal Fermi gas temperature before the sweep and using
the curve $T(T_{i})$ shown in Fig. 10, one can solve the troublesome
thermometry problem for a strongly interacting Fermi gas. From these
universal thermodynamic functions, we are also able to determine a
characterstic temperature $(T/T_{F})_{0}\simeq0.19$ or $(T_{i}/T_{F})_{0}\simeq0.16$
for a trapped Fermi gas at unitarity, which is responsible for the deviations of
thermodynamic properties away from normal Landau-Fermi-liquid behavior due to pairing effects. 
At these points, our analysis of the experimental universal function $h(\zeta)$ provides 
a new insight on the superbly precise experimental work of Nascimbène \textit{et al.} \cite{ens}.

These thorough comparisons between theory and experiment provide a
motivation for further developing the challenging many-body theory
of a strongly interacting Fermi gases. It is impressive that the simplest
Gaussian pair fluctuation approach gives such excellent agreement
with the experimental data, especially in the below threshold regime
characterized by long-range superfluid order. Yet, it fails to predict
the correct normal-superfluid transition temperature. More work is
need to understand the reason for this, but at this stage, we feel
that the GPF approximation serves as a good starting point for further
theoretical work. Recalling that the GPF approximation includes only
the two-body correlations (see, for example, Fig. 1a), a natural way
to extend this may be to consider three-body or four-body correlations,
in which three or four fermions interact with each other in the scattering
process. The thermodynamic potential may be worked out with inclusion
of all three-body or four-body scattering matrices. In this manner,
we would recover correctly the equation of state predicted by the
higher-order (i.e., $3^{nd}$ and $4^{th}$ order) quantum
virial expansion theory at high temperatures. We believe that it will
also lead to a more reasonable critical temperature and remove the
spurious bend-back structure close to $T_{c}$ as shown in the Gaussian
pair fluctuation theory.

On the other hand, the quantum virial expansion gives us another means
for theoretical development, from a very different point of view.
We have already shown the wide applicability of this expansion for
the equation of state of a strongly interacting Fermi gas in harmonic
traps, down to temperatures as low as $\sim0.4T_{F}$. We have conjectured
that it may be applicable down to the superfluid transition temperature,
with inclusion of higher-order virial coefficients. Here, we can also
develop a virial expansion for more crucial dynamical properties,
such as the dynamic structure factor and single-particle spectral
function, as measured recently at Swinburne \cite{swinexpt} and at
JILA \cite{jila3}. The quantum virial expansion may therefore solve
the troublesome problem of understanding a normal yet strongly interacting
Fermi gas, although theoretical predictions beyond third order are
not yet available.

Finally, we note that there is huge interest in determining the detailed
behavior of a strongly interacting Fermi gas near the normal-superfluid
transition. Our comparative study, based on the most recent theoretical
and experimental results, may provide useful insights for this future
research. 

\ack

We acknowledge S. Nascimbène and C. Salomon for providing us with
the experimental data prior to publication, and ENS for funding a
research visit. This research was supported by the Australian Research
Council Center of Excellence for Quantum-Atom Optics, the Australian
Research Council Discovery Project Nos. DP0984522 and DP0984637. It
was also supported by the National Natural Science Foundation of China
Grant No. 10774190, and the National Basic Research Program of China
(973 Program) Grant Nos. 2006CB921404 and 2006CB921306.


\begin{thebibliography}{58}
\bibitem{footnote} We use local thermodynamic potential to define
the universal function $h(\zeta)$. Also, $h(\zeta)$ defined in the
experimental paper \cite{ens} is renormalized by the pressure of
an ideal, single-component Fermi gas. It is therefore a factor of
2 larger than our universal function defined in Eq. (\ref{hzeta}).
\bibitem{footnote0} The sum is formally divergent at high Matsubara frequencies at $\tau=0$. To overcome the divergence, we rewrite the Green function, $G(K)=G_0(K)+\Delta G(K)$. The Matsubara frequency sum for the free Green function $G(K)$ can be calculated analytically in the correct limit of $\tau \rightarrow 0^+$ and the sum for the residue difference $\Delta G(K)$ is numerically convergent.
\bibitem{footnote2} Note that, in our earliest work \cite{hld2},
in order to calculate the trapped equation of state, we adopted a
phenomenological interpolation to remove the spurious discontinuous
structure around $T_{c}$ for the uniform NSR equation of state. For
the purpose of avoiding any adjustable parameters, this strategy is
not used in the current calculations. The interpolation strategy only
leads to small difference around $T_{c}$.
\end{thebibliography}
\end{document}